\documentclass[useAMS,usenatbib]{mnras}
\usepackage{newtxtext,newtxmath}

\usepackage[T1]{fontenc}

\DeclareRobustCommand{\VAN}[3]{#2}
\let\VANthebibliography\thebibliography
\def\thebibliography{\DeclareRobustCommand{\VAN}[3]{##3}\VANthebibliography}

\usepackage{graphicx}	
\usepackage{amsmath}	
\usepackage{verbatim}
\usepackage{pgf}
\usepackage{hyperref}
\usepackage{adjustbox}
\usepackage{caption}
\usepackage{arydshln}
\usepackage{scalerel}
\usepackage{tablefootnote}

\newcommand{\psr}{PSR~J1643$-$1224}
\newcommand{\veff}{$\textbf{v}_{\rm{eff}}$}
\newcommand{\ds}{$d_{\rm scr}$}
\newcommand{\dpsr}{$d_{\rm psr}$}
\newcommand{\dmu}{\,\mathrm{pc}\,\mathrm{cm}^{-3}}
\newcommand{\si}[1]{ {\mathrm{#1}} }

\title[Scintillation arcs of {\psr} using LEAP]{Modelling annual scintillation arc variations in {\psr} using the Large European Array for Pulsars}

\author[Geetam Mall et al.]{G. Mall$^{1,2,3}$\thanks{E-mail:gmall@mpifr-bonn.mpg.de}, R. A. Main$^{1}$,  J. Antoniadis$^{4,5,1}$, C. G. Bassa$^{6}$,  M. Burgay$^{7}$,  S. Chen$^{8,9,10}$, I. Cognard$^{10,8}$, \newauthor R. Concu$^{7}$, A. Corongiu$^{7}$, M. Gaikwad$^{1}$, H. Hu$^{1}$, G. H. Janssen$^{6,11}$, R. Karuppusamy$^{1}$,  M. Kramer$^{1,12}$,\newauthor K. J. Lee$^{13}$,   K. Liu$^{1}$,  J. W. McKee$^{2}$, A. Melis$^{7}$, M. B. Mickaliger$^{12}$, D. Perrodin$^{7}$, M. Pilia$^{7}$, \newauthor A. Possenti$^{7,14}$,  D. J. Reardon$^{15, 16}$, S. A. Sanidas$^{17,12}$,
T. Sprenger$^{1}$, B. W. Stappers$^{12}$, L. Wang$^{12, 18}$, \newauthor O. Wucknitz$^{1}$, W. W. Zhu$^{18}$ \\
$^{1}$Max-Planck-Institut f\"{u}r Radioastronomie, Auf dem H\"{u}gel 69, 53121, Bonn, Germany\\
$^{2}$Canadian Institute for Theoretical Astrophysics, University of Toronto, 60 St. George Street, Toronto, ON M5S 3H8, Canada\\
$^{3}$Department of Physics, University of Toronto, 60 St. George Street, Toronto, ON M5S 1A7, Canada\\
$^{4}$Institute of Astrophysics, FORTH, Dept. of Physics, University of Crete, Voutes, University Campus, GR-71003 Heraklion, Greece \\
$^{5}$Argelander Institut f\"{u}r Astronomie, Auf dem H\"{u}gel 71, 53121, Bonn, Germany\\
$^{6}$ASTRON, the Netherlands Institute for Radio Astronomy, Oude Hoogeveensedijk 4, 7991 PD Dwingeloo, The Netherlands \\
$^{7}$INAF - Osservatorio Astronomico di Cagliari, via della Scienza 5, I-09047 Selargius (CA), Italy \\
$^{8}$Station de radioastronomie de Nan{\c c}ay, Observatoire de Paris, PSL Research University, CNRS/INSU F-18330 Nan{\c c}ay, France \\
$^{9}$FEMTO-ST, Department of Time and Frequency, UBFC and CNRS, F-25030 Besan\c{c}on, France \\
$^{10}$Laboratoire de Physique et Chimie de l'Environnement et de l'Espace LPC2E CNRS-Universit{\'e} d'Orl{\'e}ans, F-45071, Orl{\'e}ans, France \\
$^{11}$Department of Astrophysics/IMAPP, Radboud University, P.O. Box 9010, 6500 GL Nijmegen, The Netherlands \\
$^{12}$Jodrell Bank Centre for Astrophysics, School of Physics and Astronomy, The University of Manchester, Manchester M13 9PL,UK \\
$^{13}$Kavli institute for astronomy and astrophysics, Peking University, Beijing 100871,P.R.China \\
$^{14}$ Università di Cagliari, Dipartimento di Fisica, S.P. Monserrato-Sestu Km 0,700, I-09042 Monserrato (CA), Italy \\
$^{15}$Centre for Astrophysics and Supercomputing, Swinburne University of Technology, P.O. Box 218, Hawthorn, Victoria 3122, Australia \\
$^{16}$Australian Research Council Centre of Excellence for Gravitational Wave Discovery (OzGrav) \\
$^{17}$Anton Pannekoek Institute for Astronomy, University of Amsterdam, Science Park 904, 1098 XH Amsterdam, The Netherlands \\ 
$^{18}$National Astronomical Observatories, Chinese Academy of Sciences, A20 Datun Rd, Chaoyang District, Beijing 100012, P.\,R.\,China \\
}

\date{Accepted XXX. Received YYY; in original form ZZZ}

\pubyear{2020}

\begin{document}
\label{firstpage}
\pagerange{\pageref{firstpage}--\pageref{lastpage}}
\maketitle
 
\begin{abstract}
In this work we study variations in the parabolic scintillation arcs of the binary millisecond pulsar {\psr} over five years using the Large European Array for Pulsars (LEAP). The 2D power spectrum of scintillation, called the secondary spectrum, often shows a parabolic distribution of power, where the arc curvature encodes the relative velocities and distances of the pulsar, ionised interstellar medium (IISM), and Earth. We observe a clear parabolic scintillation arc which varies in curvature throughout the year. The distribution of power in the secondary spectra are inconsistent with a single scattering screen which is fully 1D, or entirely isotropic. We fit the observed arc curvature variations with two models; an isotropic scattering screen, and a model with two independent 1D screens. We measure the distance to the scattering screen to be in the range 114-223\,pc, depending on the model, consistent with the known distance of the foreground large-diameter HII region Sh 2-27 ($112\pm17$\,pc), suggesting that it is the dominant source of scattering.
We obtain only weak constraints on the pulsar's orbital inclination and angle of periastron, since the scintillation pattern is not very sensitive to the pulsar's motion, since the screen is much closer to the Earth than the pulsar. More measurements of this kind - where scattering screens can be associated with foreground objects - will help to inform the origins and distribution of scattering screens within our galaxy.

\end{abstract}



\section{Introduction}
Pulsars are remarkably stable clocks. This property has proved them to be incredibly successful laboratories for testing the predictions of general relativity using pulsar timing \citep{taylor1979measurements, kramer2006tests}. 
Pulsar timing arrays (PTAs) aim to detect gravitational waves (GWs) by monitoring many millisecond pulsars (MSPs) over time to measure a spatially correlated signal in their timing residuals \citep{hellings+83}. The primary PTAs to date are the European Pulsar Timing Array (EPTA, \citealt{van2011placing}), which combines data from different European telescopes; the North American Nanohertz Observatory for Gravitational Waves (NANOGrav, \citealt{2013Demorest}); the Parkes Pulsar Timing Array (PPTA, \citealt{manchester2013parkes}); and the International Pulsar Timing Array (IPTA, \citealt{hobbs2010international}), which is a collaboration between the aforementioned individual PTAs. 

Recently, several PTAs reported a detection of a common red-noise process in their 12.5 year dataset \citep{arzoumanian+20, 2021+Goncharov, chen+21}, but did not observe the significant spatial correlation needed to claim a GW detection, further explored by \citet{2021+Goncharov} who argue that a signal of this type can arise from pulsars with independent red noise properties. As PTAs may be nearing detection of a GW signal, we need to better understand all sources of correlated noise in timing residuals; a large contributing factor is the ionized interstellar medium (IISM) which introduces time-varying, chromatic variations in the electron column density, and multipath propagation.

The effects of multipath propagation can often be seen through scintillation, a pattern in time and frequency caused by interference between different deflected images of the pulsar. Scintillation is now commonly studied using the secondary spectrum -- the 2D power spectrum of scintillation -- where a single dominant scattering screen results in a parabolic distribution of power (\citealt{stinebring2001faint}, \citealt{walker2004interpretation}, \citealt{cordes2006theory}). While pulsar timing is primarily sensitive to changes in radial motion, the rate of scintillation (or equivalently, the scintillation timescale) depends on the velocity on the plane of the sky. Measurements of variable scintillation rate can then be used to obtain an additional constraint on the pulsar's orbit \citep{lyne84, rickett+14, reardon+19}. 

Measurements of the scintillation timescale are dependent on the distribution of power along the scattering screen, while the curvatures of scintillation arcs are far less model dependent.
Annual and orbital variations in arc curvature can be used to measure the screen distance and geometry, and precisely measure orbital inclinations and angle of the periastron, as shown in 16 years of scintillation arc measurements of PSR J0437$-$\,4715 by \citet{reardon+2020}.

In this paper we study \psr, a $4.622\, \si{\rm{ms}}$ period pulsar in a 147\,day binary orbit with a white dwarf companion, which is observed as part of all aforementioned PTAs. We summarize the relevant theory of scintillation arcs needed for our paper in Section 2, discuss our observations and data reduction in Section 3, and interpret our arc curvature measurements in Section 4. In Section 5 we describe the models used for the arc curvature variations, and  we present our results in Section 6. Finally, we discuss the ramifications of our results in Section 7. 

\section{Background on scintillation}
Pulsar scintillation is caused by deflection of pulsar signals by inhomogeneities in the electron densities in the IISM between the pulsar and observer. These deflections create multiple images, which interfere with each other and produce an interference pattern which changes with time due to the relative velocities between the pulsar, the IISM and the Earth. The dynamic spectrum $I(f, t)$ shows the observed intensity as a function of frequency $f$ and time $t$. The squared modulus of the 2D Fourier transform of the dynamic spectrum, $I_2 = |\tilde{I}(f_D, \tau)|^2$, is called the secondary spectrum, where the $\tilde{I}$ denotes a Fourier transform. The secondary spectrum expresses the power as a function of the Doppler rate $f_{D}$ and geometric time delay $\tau$ between each pair of interfering images \citep{stinebring2001faint, walker2004interpretation, cordes2006theory}. 

A notable feature of secondary spectra are parabolic arcs (and sometimes inverted arclets), which imply the presence of a dominant and often anisotropic scattering screen between the pulsar and observer (\citealt{stinebring2001faint}, \citealt{walker2004interpretation}, \citealt{cordes2006theory}). The arc curvature $\eta$ of a parabolic arc at central observing wavelength $\lambda$ is given by
\begin{equation}
\eta = \frac{d_{\rm{eff}}\,\lambda^2}{2\, c\, \textbf{v}_{\rm{eff}}^2\, \cos^2 \alpha}, 
\label{eq:curvature}
\end{equation}
where $c$ is the speed of light, $\alpha$ is the angle between anisotropy axis of the screen on the plane of the sky and the effective velocity $\textbf{v}_{\rm{eff}}$, which depends on the velocities of the pulsar $\textbf{v}_{\rm psr}$, the IISM $\textbf{v}_{\rm{scr}}$ and the Earth $\textbf{v}_{\earth}$ perpendicular to the line of sight
\begin{equation}
\textbf{v}_{\rm{eff}} = \Big(\frac{1} {s} -1\Big) \textbf{v}_{\rm{psr}} + \textbf{v}_{\earth} - \frac{1}{s}\textbf{v}_{\rm{scr}},
\label{eq:veff}
\end{equation}
and the effective distance $d_{\rm{eff}}$ is given by
\begin{equation}
d_{\rm{eff}} = \Big(\frac{1} {s} -1\Big) d_{\rm{psr}},
\end{equation}
where $d_{\rm psr}$ and $d_{\rm {scr}}$ are the distance to the pulsar and screen, respectively, and $s \equiv 1 - d_{\rm{scr}}/d_{\rm psr} $. 
 
The variation of the observed arc curvature with time then depends on the distance, geometry and velocity of the scattering screen, as well as the distance and velocity of the pulsar. These properties will be used in Section \ref{sec:modelling}.

\section{Observations and Data}
The data description and reduction in this work are largely the same as in \citet{main+2020}; in this section we reiterate the important points, and specific reduction parameters for {\psr}.

\subsection{LEAP data}
The Large European Array for Pulsars (LEAP) is an experiment designed to increase the sensitivity of pulsar timing observations, by coherently combining signals of the five largest European telescopes. These telescopes are the Effelsberg Telescope, the Nan{\c c}ay Radio Telescope, the Sardinia Radio Telescope, the Westerbork Synthesis Radio Telescope, and the Lovell Telescope at Jodrell Bank. The data from the five telescopes are coherently added, and the resulting signal to noise (S/N) is the linear sum of the S/N from the individual telescopes \citep{ 2016MNRAS.456.2196B}. Combining these dishes results in an effective aperture equivalent a 195-m diameter circular dish.

LEAP has been observing more than twenty MSPs monthly since 2012, at a frequency band centered on 1396 MHz with a bandwidth of 128 MHz, divided into contiguous 16\,MHz sub-bands. LEAP observes with whichever telescopes are available, and the baseband data are correlated and coherently added in software at the Jodrell Bank Observatory \citep{2017A&C....19...66S}. The coherently combined baseband data are stored on magnetic tapes and can be retroactively processed to generate pulse-profile data at arbitrary time and/or frequency resolution. The high sensitivity and flexible data product have allowed LEAP to carry out more than just timing analyses, such as studies of MSP single pulses \citep{2016MNRAS.463.3239L,2019MNRAS.483.4784M} and scintillation properties \citep{main+2020}. Note that observations in 2012 had short observing times of $\sim$10\,minutes to allow for periodic scans on a phase calibrator, while from 2013 onwards the typical observing times were extended to 30-60 minutes. As such, in this work we select all observations of {\psr} from 2013 until 2018, beyond which this pulsar was no longer observed as part of the regular monthly LEAP programme.

\subsection{Creating dynamic and secondary spectra}
For each 16\,MHz sub-band, we use \textsc{dspsr} \citep{2011PASA...28....1V} to fold the coherently added baseband data into 10\,s time bins, 16 phase bins, 2048 frequency channels of width 7.8125\,kHz.
These are combined in frequency using psradd from \textsc{psrchive} \citep{Hotan+2004,2012AR&T....9..237V} to create the final folded spectrum. The small channel width of 7.8125\,kHz allows us to resolve time delays caused by scintillation up to $64 \si{\mu}s$. We sum polarizations to form total intensity, resulting in a 3-dimensional data cube with dimensions of time, frequency, and phase.

Before creating the dynamic spectrum, we flag and mask subintegrations influenced by radio frequency interference (RFI), and remove the influence of the bandpass. We sum over time and frequency to form the pulse profile, the bottom half of the profile are selected as the off-pulse region. For every sub-integration, we compute the standard deviation in the off-pulse region, and values more than $5\sigma$ greater than the mean of the data are masked. The off-pulse region, rather than the full phase window, is used so as to not inadvertently mask bright subintegrations caused by scintillation maxima. The data cube is then divided by the time average of the off-gates to remove the effects of the bandpass. The cleaned time and frequency averaged pulse profile is used as a template. We use the template to weight each phase bin, then sum over phase to create the dynamic spectrum $I(t,f)$.

The arc curvature changes as a function of frequency (see Eqn \ref{eq:curvature}); approaches to deal with this include Fourier transforming over $\lambda$ instead of frequency \citep{reardon+2020} or over a time axis scaled by frequency \citep{sprenger+2021}. In our case, the fractional bandwidth is small, so we compute the secondary spectrum $\big|\tilde{I}(f_D, \tau)\big|^2$ directly as the squared amplitude of the 2-dimensional Fourier transform of the dynamic spectrum $I(t,f)$.

\begin{figure*}
\centering
\includegraphics[width=1.0\textwidth, trim=3cm 5.5cm 2cm 5.0cm, clip=true]{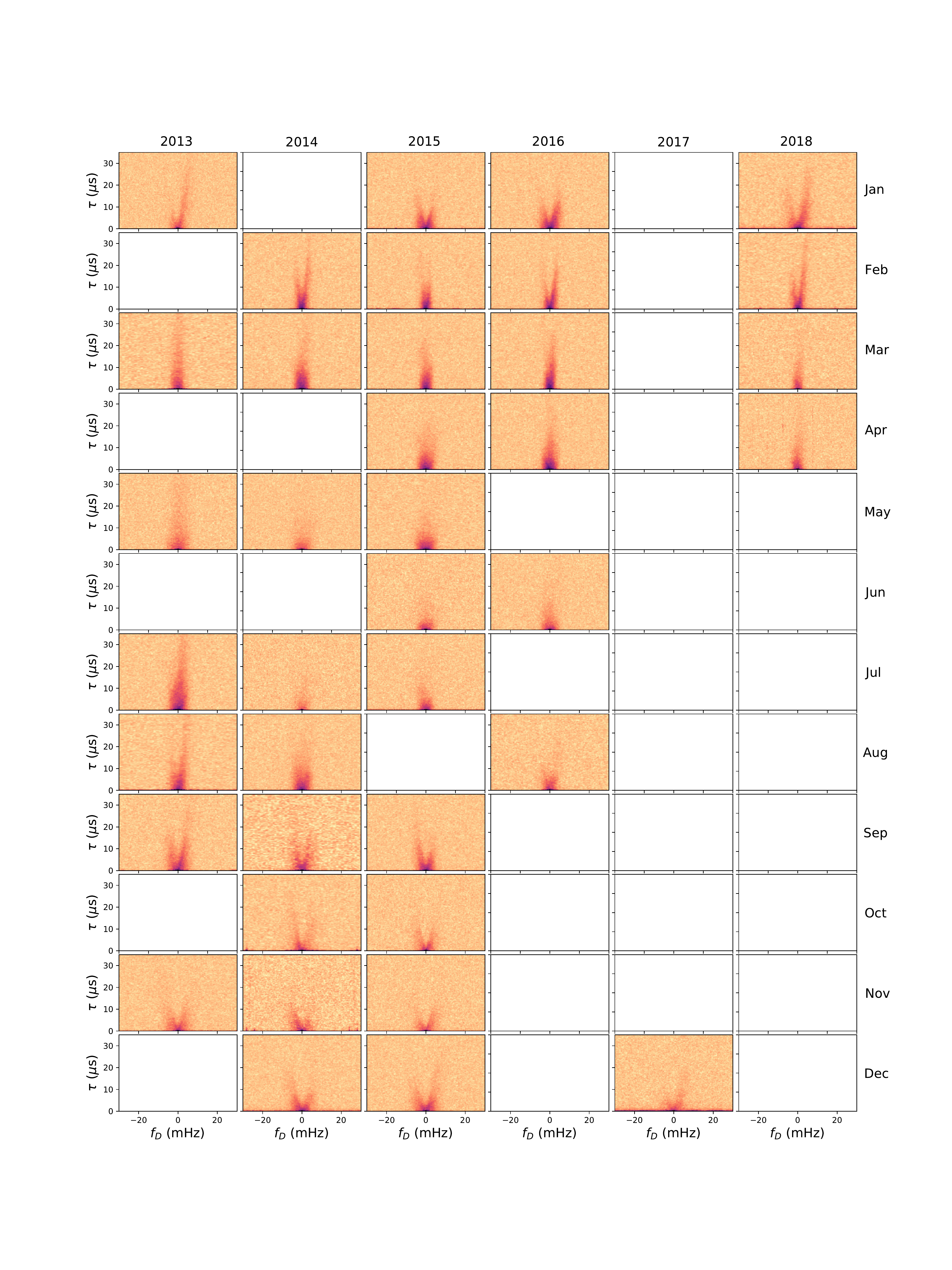}\\
\caption{Secondary spectra showing parabolic arcs for all of observations of {\psr} used in this paper. 
Rows correspond to months (Jan--Dec) and columns correspond to years (2013--2018). The blank entries in our data correspond to missing observations. We can see variations in arc curvature by comparing the parabolae. This variation is due to annual and orbital variations from the velocity of Earth and the pulsar respectively, as descried in section \ref{sec:arc_qualitative}}
    \label{fig:panorama}
\end{figure*}   

\section{Secondary Spectra}

\subsection{Interpretation of the observed arcs}
\label{sec:arc_qualitative}
Figure \ref{fig:panorama} shows the secondary spectra created for each of our observations, obtained using the methods described in Section 3.3. We see clear scintillation arcs, varying annually. Orbital variation would be clearly observable by comparing arcs on the same month across several years; the 147 day pulsar orbital period is coincidentally close to a 2/5 of a year, effectively causing a 2:5 orbital resonance. After one year, the pulsar will have moved over 2.5 orbits, while after two years the pulsar will have moved over 5 orbits, and the orbital motion and Earth's motion will then be aligned and anti-aligned on alternating years. However, we do not see a strong effect of the orbital motion in our data, with the arc curvature of a given month appearing similar at every year.

An immediately apparent feature of the secondary spectra is how their distribution of power varies throughout the year. Months September -- February show a clear arc indicative of a dominant anisotropic scattering screen. However, months March -- July do not show discernible arcs, but rather a more diffuse distribution of power across $f_{D}$. In these months, it often appears that the power at low and high time delays follow parabolae of different curvatures.  

The observed secondary spectra cannot be explained either through a single 1D screen, or a fully isotropic screen. A 1D screen may explain the clean arcs, but would collapse to a line on $f_{D}=0$ when $\textbf{v}_{\rm{eff}}^2\, \cos^2 \alpha = 0$, while an isotropic screen would likely not show such clear arcs, and would have a qualitatively similar distribution of power when seen at different angles of $\textbf{v}_{\rm eff}$. To fully explain our observations, we need either a second misaligned screen, or for the dominant screen to be elongated with an axial ratio $\gtrsim 2$.
We consider both of the possible models in our analysis in the later sections. The possible screen geometries, and their resultant secondary spectra are explored further using simple illustrative simulations in Appendix \ref{sec:screengeometry}.

\subsection{Measuring arc curvatures}
\label{sec:curvature}

To measure arc curvatures, we first average the secondary spectrum to 256 sub-samples of width $0.5\,\si{\mu}s$ along the $\tau$-axis. For each $\tau$ sub-sample we then fit a double-peaked Gaussian curve as
\begin{equation}
\log_{10}\left(|I(f_{D}, \tau_{i})|^{2} \right) = 
a \exp{ \left(\frac{-(f_{D}-\mu_{f_{D},i})^2}{2\sigma^2} \right) } + b \exp{ \left(\frac{-(f_{D}+\mu_{f_{D},i})^2}{2\sigma^2} \right) }.
\end{equation}
While not a physically motivated choice, this approximation is useful for finding the power centroid for each $\tau_i$, used to fit for arc curvatures.
We take only values with $a$ or $b$ greater than 4 times the RMS of the background noise for each $I(f_{D}, \tau_{i})$, and remove points which converged to an anomalously large ($\sigma > 5\,$mHz) or small ($\sigma < 0.1\,$mHz) Gaussian width. We then have a series of independent data points $\tau_{i}$, and dependent data points $\mu_{f_{Di}}$ with uncertainties, which we fit with $|f_{D}| = \sqrt{\tau / \eta}$. The proportionality constant $1/\sqrt{\eta}$ from the square root fitting is directly proportional to $|v_{\rm{eff}}|$. 

\begin{figure*}
 \centering
   \includegraphics[width=1.0\columnwidth, trim=0cm 1cm 2cm 2cm, clip=true]{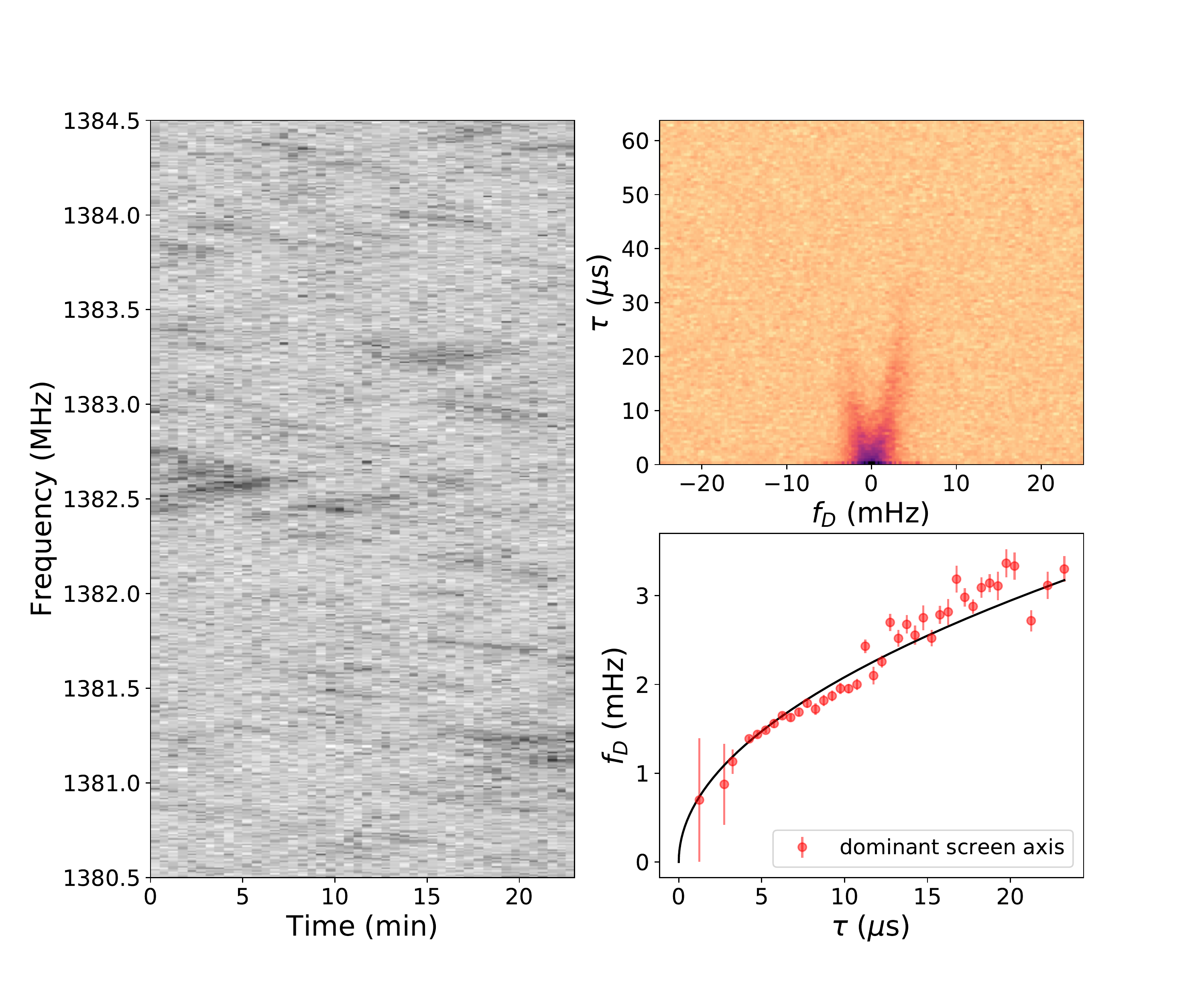}
   \includegraphics[width=1.0\columnwidth, trim=0cm 1cm 2cm 2cm, clip=true]{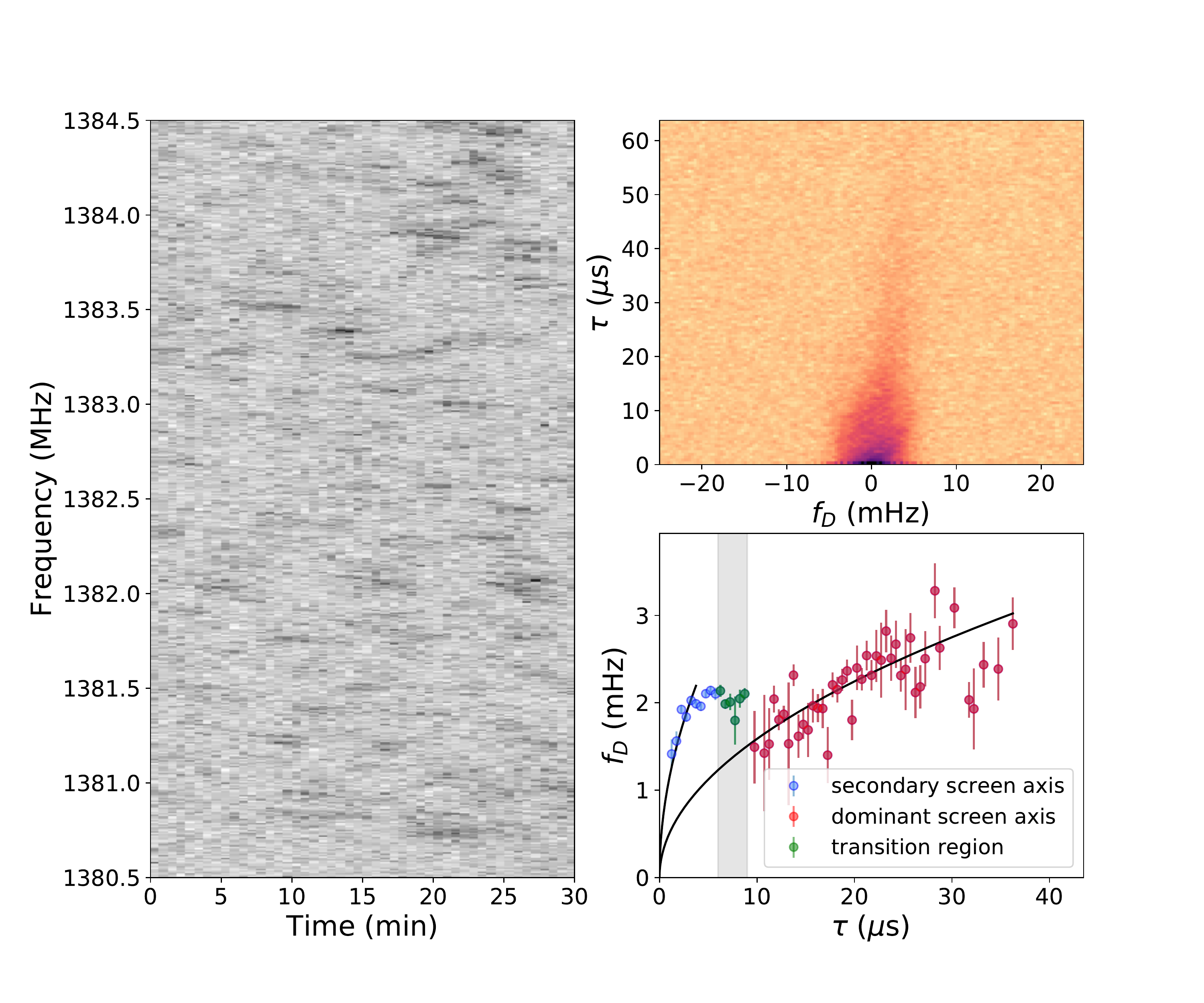}\\
   \caption{Dynamic and secondary spectra, with curvature fits for February 2014 \textit{(left)} and July 2013 \textit{(right)} observations. In each panel: \textit{Left images}: Dynamic spectra  zoomed into a bandwidth of 4\,MHz to discern the fine-scale of scintillation. Dynamic spectra were binned to 30\,s for plotting purposes. \textit{Top right images}: Corresponding secondary spectra, with same scaling as in Figure \ref{fig:panorama}. \textit{Bottom right subplots:} Fitted points $\mu_{f_D,i}$ in slices $\tau_{i}$, as described in Section \ref{sec:curvature}, from which $\eta$ is measured. 
   February 2014 shows a distinct arc well described by one screen and associated diagonal striping in the dynamic spectra. In comparison, July 2013 shows power distributed in a way inconsistent with a single 1D or isotropic screen, as can be seen in the different behaviour at low and high $\tau$.}
  \label{fig:fuzzyarc}
  \end{figure*}

As described in Section \ref{sec:arc_qualitative}, at certain months of the year we see wide and diffuse arcs at low time delays, which cannot be explained by a single 1D or isotropic screen.  
In such cases, the secondary spectrum may not follow a single parabola.
  To account for the presence of a second screen, or a secondary screen axis, we restrict our fit to low time delays (taken as $\tau \lesssim 6 \mu$\,s ). The wide parabola at low time delays represents the screen with the highest projected velocity - either a second scattering screen, or the points along the axis of motion in a 2D screen. Attempting to measure a curvature of the points at high time delays is more difficult, and may lead to a biased measurement depending on the screen model.  An example fit is shown in Figure \ref{fig:fuzzyarc}, showing a case where there is a clear and dominant arc (top panels), and a case where the curvature at low time delays does not match the curvature at high time delays (bottom panels). The transition between the behaviour at low and high values of $\tau$ happens at $\sim 6 - 9\,\si{\mu}s$, motivating our choice of the $\tau$ cutoff in our fitting.

\section{Modelling the varying arc curvature}
\label{sec:modelling}
From Equations \ref{eq:curvature} and \ref{eq:veff}, the changing velocity from the Earth's and pulsar's orbit results in arc curvature variations. A model of the arc curvature will include properties of the pulsar, specifically the distance and proper motion, which are already constrained through pulsar timing, and unknown values of orbital inclination ($i$) and longitude of ascending node ($\Omega$). 
We take measurements of the pulsar's distance ( \dpsr $= 0.85 \,\pm 0.35 \,$ kpc)
from the EPTA timing results of \citet{desvignes2016high}. During modelling we fix proper motions to their timing values, but allow distance to vary, using the timing value as a Gaussian prior.

The distance measurement \dpsr $= 0.85 \,\pm \,0.35 \,$ kpc comes directly from the parallax measurement ($\pi = 1.17 \, \pm \, 0.26 \,$ mas from \citealt{desvignes2016high}), and is consistent with the recent PPTA distance \citep{reardon+21} of $\pi = 0.82 \, \pm \, 0.17 \,$ mas, \dpsr $= 1.2^{0.4}_{-0.3} \,$ kpc.  The quoted distance from \citet{desvignes2016high} is \dpsr $= 0.76 \,\pm \,0.16 \,$ kpc, which applies the Lutz-Kelker bias correction including a constraint from the pulsar's luminosity estimate.  This reduces the error by more than a factor of 2 compared to the direct parallax distance - we adopt the above value and errors to be conservative. The proper motion values are $\mu_{\alpha}= 6.04\,\pm 0.04\, \rm mas \, yr ^{-1}, \mu_{\delta} = 4.07\,\pm 0.15 \, \rm mas \, yr ^{-1}$. 

For a given screen model, we must also include the screen distance (\ds), and parameters related to the screen velocities and geometry, which we explicitly describe in the following sections.
The function to compute arc curvatures from the the pulsar's orbital motion and Earth's known velocity was taken from \textsc{scintools}\footnote{\url{https://github.com/danielreardon/scintools/}} described in \citet{reardon+2020}, slightly modified to fit for ${1/\sqrt{\eta}}\propto v_{\rm eff}$ instead of fitting $\eta$ directly, and adding a two-screen model as described below. Our Markov chain Monte Carlo (MCMC) fits are performed using the \textsc{emcee} implementation in \textsc{lmfit} \citep{2014zndo.....11813N}.

\subsection{Isotropic model}
\label{sec:isotropic}
In an isotropic screen, $\cos^{2} \alpha = 1$ for any orientation of \veff, and the arc curvature depends only on the magnitude of \veff\ on the plane of the sky. A full model of the screen needs only three parameters, the screen distance, and the 2D screen velocity on the plane of the sky.

As discussed in Section \ref{sec:arc_qualitative} and Appendix \ref{sec:screengeometry}, the observed secondary spectra cannot be fully described by a single isotropic screen, needing either multiple screens or an elongated screen with axial ratio $\gtrsim 2$. However, in measuring the arc curvatures at low enough time delays, we are always measuring the magnitude of \veff\ on the plane of the sky, and the model for the arc curvature variations of any 2D screen is equivalent to the isotropic case. This was one of the primary motivations for restricting time delays to $\tau < 6\,\mu$s, below the visible transition. A fit to the full power distribution of the secondary spectrum for an elongated 2D screen would require at least 2 more parameters: the axial ratio and orientation of the screen.

\subsection{Anisotropic two-screen model}
\label{sec:anisotropic}
For the case of a 1D screen model, it is only possible to measure {\veff} parallel to the screen.
The observed secondary spectra can be qualitatively produced through the existence of two screens (Section \ref{sec:arc_qualitative} and Appendix \ref{sec:screengeometry}). In this model, we have two screens, each with a separate distance $d_{\rm{s1}}$ and $d_{\rm{s2}}$, angle  ${\Psi_1}$ and $\Psi_2$, and velocity along the screen's axis $V_{\rm{s1},\Psi_{1}}$ and $V_{\rm{s2},\Psi_{2}}$.  

The measured curvature at low time delays is a measure of the screen with maximum projected effective velocity at any given time (or more precisely, the screen with minimum arc curvature at any given time). For any set of model parameters, there are model predictions of $\eta$ arising from both screens, our final model which is fit takes the minimum curvature between these models at any given time.
   
\subsection{Treatment of Uncertainties}

In Section \ref{sec:curvature}, we described our measurements of the arc curvature $\eta$ and the formal statistical uncertainties $\delta \eta$. However, the formal uncertainties $\delta \eta$ may be underestimated due to unmodelled systematic errors, which could arise from, e.g., asymmetric power distribution of the arcs, unresolved arclets, or contribution to the arc curvature from dimension perpendicular to the primary screen axis. Underestimated errors will lead to biases in the final posterior distributions.  To address this issue, we take the approach of using `EQUAD' and `EFAC' values typically used in pulsar timing, which describe the corrected errors as 
\begin{equation}
\delta \eta_{\rm corr} = \sqrt{ (\rm{EFAC\, \times\,\delta \eta})^{2} + \rm{EQUAD}^{2} }.
\end{equation}
In a grid of EQUAD, EFAC values (EQUAD ranging from  $0-2$, EFAC ranging from $0.5-5$ in 100 steps), we perform a KS-test on the scaled residuals to test how well they are described by a Standard Normal Distribution. We find a maximum value at $\rm{EFAC}=3.0, \rm{EQUAD}=0.36$; we adopt these values to correct the errors on $1/\sqrt{\eta}$ before performing MCMC fits.

\begin{figure*}
\centering
   \includegraphics[width=1.0\columnwidth, trim=1.5cm 1cm 2cm 2.3cm, clip=true]{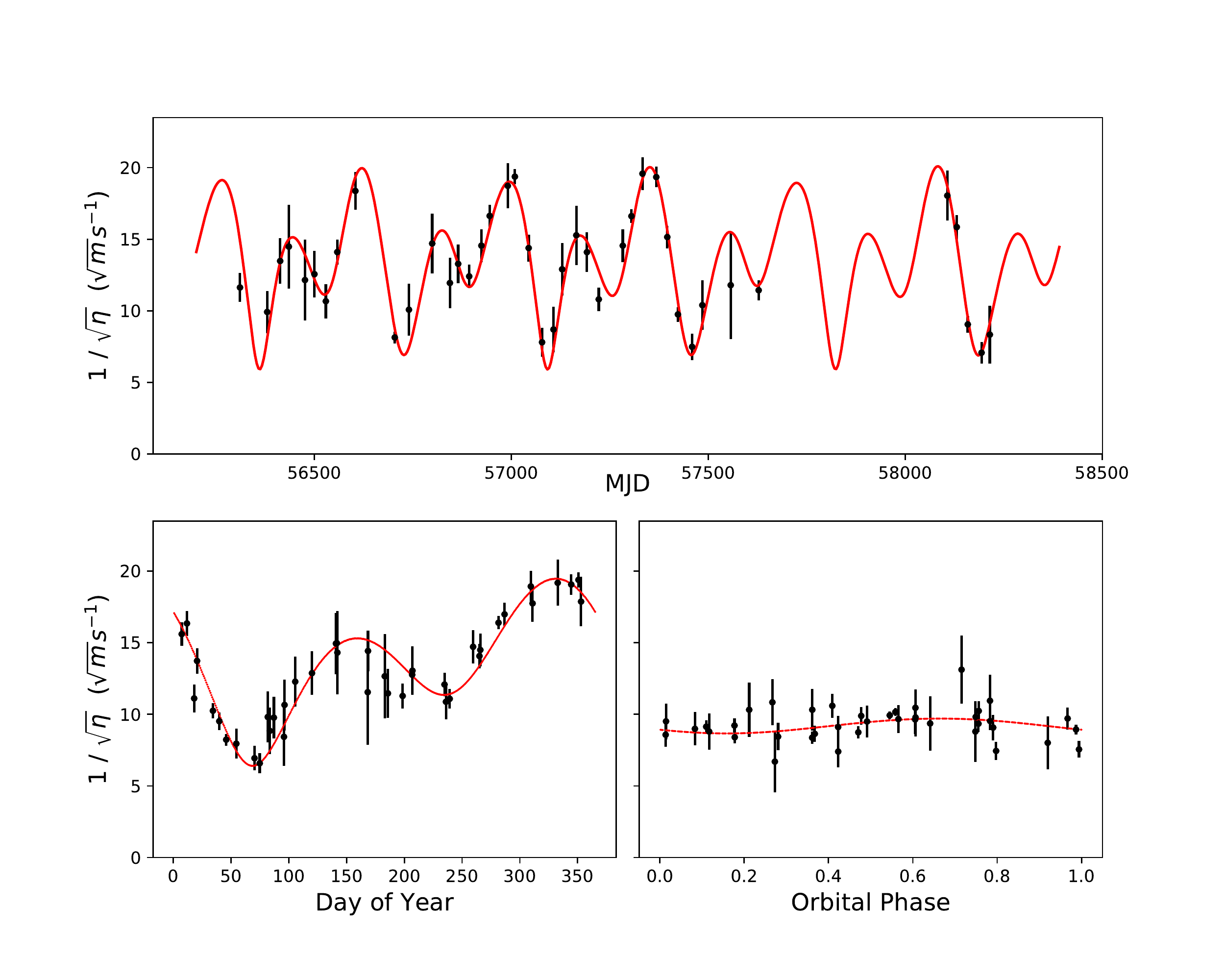}
   \includegraphics[width=1.0\columnwidth, trim=1.5cm 1cm 2cm 2.3cm, clip=true]{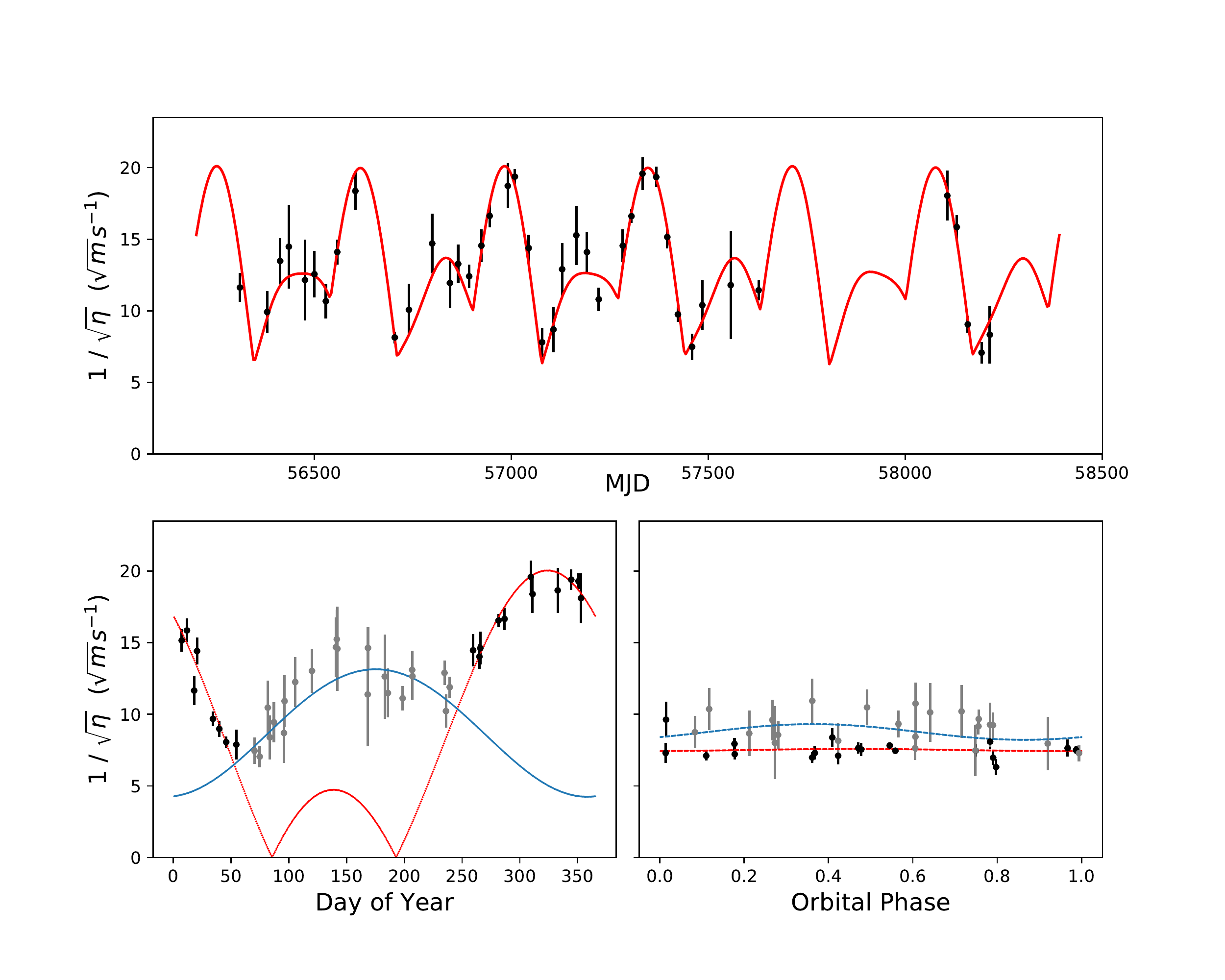}\\
    \caption{ Measurements and modelling of the variable arc curvature of \psr{}.
    The left plots show the isotropic 2D screen model, and the right show the model with two independent 1D screens, described in section \ref{sec:modelling}). Top panels show $1/\sqrt{\eta}$ vs. time, with data points in black, and model fit in red.  The below subplots show variation of $1/\sqrt{\eta}$ with time of year (left), and variation with orbital phase (right). The data points and model corresponding the the second screen are shown in grey and blue, respectively. }
    \label{fig:arcvstime}
\end{figure*}

\section{Discussion}

\subsection{Results of Model Fitting}

The best-fit parameters to the variations in arc curvature are listed in Table \ref{tab:resultstable}, and the data alongside the model fits are shown in Figure \ref{fig:arcvstime}. The bottom panels show the isolated effects of the annual and orbital variation,  after subtracting the model velocity of either the pulsar and the Earth from the data. Figures \ref{fig:mcmc2D} and \ref{fig:mcmctwoscreen} in Appendix \ref{sec:posteriors} show the full posterior probability distributions for the two models.

The isotropic and anisotropic models yield $\chi^2$ values of $\chi^2= 45.9$ and $\chi^2=37.3$ respectively, suggesting the anisotropic model is a better fit to the data. However, the anisotropic model has more free parameters, so we compute the Bayesian information criterion (BIC) for both, resulting in BIC values of 23.2 and 25.9 for the isotropic and anisotropic models respectively - by this criterion, the isotropic model is slightly preferred to describe the variations in arc curvature.

The scattering screens in both models lie closer to the Earth than the pulsar, so the effective velocity $\rm{v_{eff}}$ of the system is more sensitive to the motion of the Earth than to that of the pulsar, resulting in relatively poor constraints on orbital parameters. Despite this, the isotropic model clearly gives $i > 90^{\circ}$, resolving the sense of the orbit, while the anisotropic two-screen model finds two local solutions for $i$ and $\Omega$. In addition, the values of $i$ and $\Omega$ are consistent with recent PPTA constraints from the annual-orbital parallax of \psr{} (\citealt{reardon+21}, Fig. 3). As can be seen in Figures \ref{fig:mcmc2D} and \ref{fig:mcmctwoscreen}, we observe strong degeneracies between pulsar distance \dpsr{} and screen parameters, so obtaining accurate measurements is highly dependent on the accuracy of the priors on distance. As there has been slight tension between PTA measurements, with the previous IPTA value appearing lower \citep{verbiest2016international}, and the recent PPTA value appearing higher \citep{reardon+21} than the EPTA value \citep{desvignes2016high}, we adopted a conservative distance prior, as described in Section \ref{sec:modelling}.

\begin{table}
    \caption{The main table shows the resulting values from the fit, $\chi^{2}$, degrees of freedom (dof), and Bayesian Information Criterion (BIC) values for both models.The parameters are defined in Section \ref{sec:modelling}} .
    
    \begin{tabular}{ccc}
    \hline
    Parameters & Isotropic & Anisotropic \\
    \hline
    \dpsr\,(kpc) & $1.0 \pm 0.3$ & 
    $1.2 \pm 0.3^{\text{*}}$ \\
    
    \ds\,(kpc) & $0.2$ & 
    ...   \\
    
    $V_{\rm{scr},\alpha}$ (km s$^{-1}$) &  
    $7.6 \pm 0.8$ & 
    ... \\
    
    $V_{\rm{scr},\delta}$ (km s$^{-1}$) & 
    $-11 \pm 1$ & 
    ...  \\
    
    $d_{\rm{s1}}$ (kpc) & 
    ... & 
    $0.13$\\
    
    $V_{\rm{s1},\Psi_{1}}$ (km s$^{-1}$) & 
    ... & 
    $10 \pm 1$\\
    
    $\Psi_{1}(^{\circ}$) & 
    ... & 
    $150 \pm 4$  \\
    
    $d_{\rm{s2}}$\,(kpc) & 
    ... & 
     $0.3 \pm 0.1$  \\
    
    ${V_{\rm{s2},\Psi_{2}}}$ (km s$^{-1}$) & 
    ... & 
    $-6 \pm 3$  \\
    
    $\Psi_{2}(^{\circ}$) &
    ... & 
    $31 \pm 9$ \\
    
    $i (^{\circ}$)   
    & $135 \pm 19$
    & $95 \pm 32 \dagger$ \\
    
    $\Omega (^{\circ}$) &
    $278 \pm 17$  & $231 \pm 109 {\dagger}$\\ 
    
    $\chi^{2}$ & 45.9 &  37.3 \\
    
    $\rm{dof}$ & 35 & 32 \\
    
    BIC & 23.2 & 25.9  \\
    \hline
     \end{tabular}
     \label{tab:resultstable}
     \newline
\footnotesize{$^{*}$:\, As mentioned in Section \ref{sec:modelling}, we take a Gaussian prior of \dpsr$= 0.85\pm0.35$\,kpc.  $\dagger$:\, For the anisotropic model the solution for i and $\Omega$ converges to two local minima at i$=68^{\circ} \pm\, 16^{\circ}$, $\Omega =101^{\circ} \pm\, 44^{\circ}$ and i$=117^{\circ}\,\pm 17^{\circ}$, $\Omega= 284^{\circ}\,\pm 43^{\circ}$ (see Figure \ref{fig:mcmctwoscreen})}
\end{table}

\subsection{Screen association with Sh 2-27}
The HII region Sh 2-27, behind which {\psr} lies, is ionised by the O-star $\zeta$ Oph at a distance of $112\pm3\,\rm{pc}$ \citep{ocker+20}, and has an inferred diameter of $0.034\,\rm{kpc}$ assuming spherical symmetry \citep{harvey+2011}. The distance to the screen is then $0.112\pm0.003\pm0.017\,\rm{kpc}$, with uncertainties of the distance and the HII region's radius.

Sh 2-27 is commonly assumed to be the major contributing region for the scattering of {\psr}. In addition, several features have been observed in {\psr} which may be attributed to propagation. A variation in the flux density of {\psr} was observed from 1997 to 2000, interpreted as an extreme scattering event (ESE) and modelled as arising from a single ionizing cloud \citep{maitia+03}. Between 2010 and 2016 timing residuals for {\psr} were measured by \citet{shannon+2016} using the Parkes radio telescope, and in 2015 the pulsar displayed a timing event when a new component of emission suddenly appeared in its pulse profile. The L-band ($1-2\,$GHz) pulse profile for March 2015 showed significant TOA shifts ($\approx 10\si{\mu}s$), with larger deviations at $3\,$GHz, and no observed shift at $600\,$MHz. Due to the opposite expected scaling of the time delays with frequency, \citet{shannon+2016} concluded this event to be of magnetospheric in origin, intrinsic to the pulsar. However, a following study by NANOGrav suggests that the pulse variation may be caused by propagation, given the qualitative resemblance of the additional profile component with an echo \citep{brook+18}. 
We however do not notice any significant, qualitative change in the secondary spectra spanning this period (See Figure \ref{fig:panorama}). A quantitative analysis of the scattering time delays measured for this pulsar from LEAP data will be addressed in a future paper.

In our isotropic model, we measured the distance to the scattering screen to be $d_{\rm{scr}}$= $0.208 \pm 0.015$\,kpc, while in our two-screen anisotropic model, we measured the distance to the dominant scattering screen of $d_{\rm{s1}} = 0.129 \pm 0.015$\,kpc,  with the secondary screen at $d_{\rm{s2}}$ = $0.34 \pm 0.09$\,kpc. Depending on our choice of modelling, the scattering screen is consistent with being entirely within (in the anisotropic case), or near the boundary of (isotropic case) Sh 2-27, strongly suggesting that the scattering resulting in the observed scintillation arcs is associated with Sh 2-27.

\citet{harvey+2011} estimated the maximum possible mean electron density inside Sh 2-27 to be $\rm{n_{e,HII} = 10.1\,\pm{1.1}\,cm^{-3}}$. Using this value, \citet{ocker+20} estimated the DM contribution of Sh 2-27 to the pulsar's total observed DM to be between $\rm{DM_{HII} = 34.4\,\pm {4.5}\dmu}$ and $56.2\,\pm0.4 \dmu$, comprising at least half of the total $\rm{DM}\approx62.4\dmu$ (the range is due to slight tension between the cited IPTA parallax \citep{verbiest2016international} compared to the EPTA \citep{desvignes2016high} and PPTA \citep{reardon+2016} parallax measurements).
Given the large electron density, scattering within Sh 2-27 is almost inevitable, and the observed secondary spectra may even be the result of several scattering screens within the HII region.

We note that other studies have found associations with foreground sources. The most clear associations are for pulsars within supernova remnants - the screen in the Crab nebula has be used to probe pulsar separation of giant pulse emission regions \citep{main+21}, and \citet{yao+21} measured the distance between PSR J0538+2817 and the supernova remnant shell using scintillation arcs, suggesting a 3D spin-velocity alignment of the pulsar.
\citet{dexter+17} used the VLBA $+$ VLA to measure angular broadening of several pulsars, and were able to associate scattering screens of three sources with foreground HII regions, and three sources within Carina-Sagittarius spiral arm of the Milky Way. \citet{gupta1994refractive} found a persistent slope in the ACF of scintillation of PSR B1642$-$03, which they postulate arises from the edge of Sh 2-27 (the same HII region studied in this paper). \citet{bhat2002} used scintillation from a sample of pulsars, finding enhanced scattering in many pulsars, likely caused by the Loop I bubble.
\citet{reardon+2020} measured the distance to two scattering screens towards PSR J0437$-$4715 which could be near the edges of the Local Bubble.

\section{Conclusions}
We presented and modelled 5 years of variable scintillation arcs of \psr. Depending on the choice of screen model, the distance to the dominant scattering screen $d_{\rm scr}$ is found to be within $\sim 120-200\,$\,pc, likely associated with the foreground HII region Sh 2-27. We additionally measure an orbital inclination and angle of periastron which are consistent between our two models, but poorly constrained because the scattering screen is much closer to the Earth than to the pulsar. Generically, if we can associate scattering screens to known astrophysical objects -- particularly sources less extended than Sh 2-27 -- then scintillation arc modelling may allow for an independent determination of precise pulsar distances. 

We find that both an isotropic screen model and the two-screen model can reproduce our observed variable arc curvature, with a slight preference to the isotropic model, given its fewer free parameters. However, the appearance of the power distributions in our secondary spectra suggest that the scattering screen cannot be fully isotropic (see Appendix \ref{sec:screengeometry} for more details). Therefore to fully describe this scattering screen, we would need at least two more parameters: its degree of anisotropy and its orientation. A full modelling of the observed secondary spectra may distinguish between models, as the data are not uniquely described by a single measurement of an arc curvature. 
Additionally, sensitive future observations of {\psr} at higher frequencies could reveal multiple, sharper arcs, which could be used to inform the choice of model. 
Secondary spectra show well defined arc at higher frequencies because the thickness of the arc is strongly dependent on frequency \citep{2019+Stinebring}. 

If the scattering of a pulsar is dominated by a single, highly anisotropic scattering screen, we may be able to employ the $\theta-\theta$ method which transforms the secondary spectra variables $\tau$ and $f_D$ to angular coordinates and shows the secondary spectra with parallel linear features which are more convenient to interpret than parabolic arcs. \citet{sprenger+2021} introduces and describes the transformation and \citet{baker+21} show how it can be used to measure precise arc curvatures.
However, when multiple or two-dimensional scattering screens are present, as is the case for our observations, there may be biases, as $\theta-\theta$ assumes highly anisotropic scattering screens. The potential biases will first need to be explored in detail using simulations.

Many of our observations of secondary spectra display clear asymmetry in their power distributions. While not investigated here, such behaviour may be related to a potential local gradient in DM in the scattering screen along the direction of {\veff} (eg. \citealt{rickett+14}). In addition, the secondary spectrum shows the geometric time delay between interfering images, and under some assumptions can be used to estimate the total time delay due to multipath propagation, which may improve timing \citep{hemberger+2008,main+2020}. An investigation of the link between DM, scattering timescale, and scintillation arc variability for {\psr}, along with other LEAP sources, will be the focus of future work.

\section*{Acknowledgements}
We thank the anonymous referee for their comments which improved this work. GM thanks C. R. H. Walker and Rik van Lieshout for detailed and valuable suggestions. This work was supported by the ERC Advanced Grant ``LEAP'', Grant Agreement Number 227947 (PI M.\,Kramer). JA acknowledges support by the Stavros Niarchos Foundation (SNF) and the Hellenic Foundation for Research and Innovation (H.F.R.I.) under the 2nd Call of ``Science and Society'' Action Always strive for excellence – ``Theodoros Papazoglou'' (Project Number: 01431). KL is supported by the European Research Council for the ERC Synergy Grant BlackHoleCam under contract no. 610058. J. W. McKee is a CITA Postdoctoral Fellow: This work was supported by the Natural Sciences and Engineering Research Council of Canada (NSERC), [funding reference \#CITA 490888-16]. TS is a member of the International Max Planck Research School for Astronomy and Astrophysics at the Universities of Bonn and Cologne.

The European Pulsar Timing Array (EPTA) is a collaboration between European Institutes, namely ASTRON (NL), INAF/Osservatorio Astronomico di Cagliari (IT), the Max-Planck-Institut f{\"u}r Radioastronomie (GER), Nan{\c c}ay/Paris Observatory (FRA), The University of Manchester (UK), The University of Birmingham (UK), The University of Cambridge (UK), and The University of Bielefeld (GER), with an aim to provide high-precision pulsar timing to work towards the direct detection of low-frequency gravitational waves. The Effelsberg 100-m telescope is operated by the Max-Planck-Institut f{\"u}r Radioastronomie. Pulsar research at the Jodrell Bank Centre for Astrophysics and the observations using the Lovell Telescope are supported by a consolidated grant from the STFC in the UK. The Westerbork Synthesis Radio Telescope is operated by the Netherlands Foundation for Radio Astronomy, ASTRON, with support from NWO. 
The Nan{\c c}ay Radio Observatory is operated by the Paris Observatory, associated with the French Centre National de la Recherche Scientifique. The Sardinia Radio Telescope (SRT) is funded by the Department of Universities and Research (MIUR), the Italian Space Agency (ASI), and the Autonomous Region of Sardinia (RAS), and is operated as a National Facility by the National Institute for Astrophysics (INAF).

\section*{DATA AVAILABILITY}
The timing data used in this article shall be shared on reasonable request to the corresponding author.

\bibliographystyle{mnras}
\bibliography{pulsar} 

\begin{thebibliography}{}
\makeatletter
\relax
\def\mn@urlcharsother{\let\do\@makeother \do\$\do\&\do\#\do\^\do\_\do\%\do\~}
\def\mn@doi{\begingroup\mn@urlcharsother \@ifnextchar [ {\mn@doi@}
  {\mn@doi@[]}}
\def\mn@doi@[#1]#2{\def\@tempa{#1}\ifx\@tempa\@empty \href
  {http://dx.doi.org/#2} {doi:#2}\else \href {http://dx.doi.org/#2} {#1}\fi
  \endgroup}
\def\mn@eprint#1#2{\mn@eprint@#1:#2::\@nil}
\def\mn@eprint@arXiv#1{\href {http://arxiv.org/abs/#1} {{\tt arXiv:#1}}}
\def\mn@eprint@dblp#1{\href {http://dblp.uni-trier.de/rec/bibtex/#1.xml}
  {dblp:#1}}
\def\mn@eprint@#1:#2:#3:#4\@nil{\def\@tempa {#1}\def\@tempb {#2}\def\@tempc
  {#3}\ifx \@tempc \@empty \let \@tempc \@tempb \let \@tempb \@tempa \fi \ifx
  \@tempb \@empty \def\@tempb {arXiv}\fi \@ifundefined
  {mn@eprint@\@tempb}{\@tempb:\@tempc}{\expandafter \expandafter \csname
  mn@eprint@\@tempb\endcsname \expandafter{\@tempc}}}

\bibitem[\protect\citeauthoryear{{Arzoumanian} et~al.,}{{Arzoumanian}
  et~al.}{2020}]{arzoumanian+20}
{Arzoumanian} Z.,  et~al., 2020, \mn@doi [\apjl] {10.3847/2041-8213/abd401},
  \href {https://ui.adsabs.harvard.edu/abs/2020ApJ...905L..34A} {905, L34}

\bibitem[\protect\citeauthoryear{{Baker}, {Brisken}, {van Kerkwijk}, {Main},
  {Pen}, {Sprenger}  \& {Wucknitz}}{{Baker} et~al.}{2021}]{baker+21}
{Baker} D.,  {Brisken} W.,  {van Kerkwijk} M.~H.,  {Main} R.,  {Pen} U.-L.,
  {Sprenger} T.,   {Wucknitz} O.,  2021, arXiv e-prints, \href
  {https://ui.adsabs.harvard.edu/abs/2021arXiv210104646B} {p. arXiv:2101.04646}

\bibitem[\protect\citeauthoryear{{Bassa} et~al.,}{{Bassa}
  et~al.}{2016}]{2016MNRAS.456.2196B}
{Bassa} C.~G.,  et~al., 2016, \mn@doi [\mnras] {10.1093/mnras/stv2755}, \href
  {https://ui.adsabs.harvard.edu/abs/2016MNRAS.456.2196B} {456, 2196}

\bibitem[\protect\citeauthoryear{Bhat \& Gupta}{Bhat \& Gupta}{2002}]{bhat2002}
Bhat N.~R.,  Gupta Y.,  2002, The Astrophysical Journal, 567, 342

\bibitem[\protect\citeauthoryear{{Brook} et~al.,}{{Brook}
  et~al.}{2018}]{brook+18}
{Brook} P.~R.,  et~al., 2018, \mn@doi [\apj] {10.3847/1538-4357/aae9e3}, \href
  {https://ui.adsabs.harvard.edu/abs/2018ApJ...868..122B} {868, 122}

\bibitem[\protect\citeauthoryear{{Chen} et~al.,}{{Chen} et~al.}{2021}]{chen+21}
{Chen} S.,  et~al., 2021, \mn@doi [\mnras] {10.1093/mnras/stab2833}, \href
  {https://ui.adsabs.harvard.edu/abs/2021MNRAS.508.4970C} {508, 4970}

\bibitem[\protect\citeauthoryear{Cordes, Rickett, Stinebring  \& Coles}{Cordes
  et~al.}{2006}]{cordes2006theory}
Cordes J.~M.,  Rickett B.~J.,  Stinebring D.~R.,   Coles W.~A.,  2006, The
  Astrophysical Journal, 637, 346

\bibitem[\protect\citeauthoryear{{Demorest} et~al.,}{{Demorest}
  et~al.}{2013}]{2013Demorest}
{Demorest} P.~B.,  et~al., 2013, \mn@doi [\apj] {10.1088/0004-637X/762/2/94},
  \href {https://ui.adsabs.harvard.edu/abs/2013ApJ...762...94D} {762, 94}

\bibitem[\protect\citeauthoryear{Desvignes et~al.,}{Desvignes
  et~al.}{2016}]{desvignes2016high}
Desvignes G.,  et~al., 2016, Monthly Notices of the Royal Astronomical Society,
  458, 3341

\bibitem[\protect\citeauthoryear{{Dexter} et~al.,}{{Dexter}
  et~al.}{2017}]{dexter+17}
{Dexter} J.,  et~al., 2017, \mn@doi [\mnras] {10.1093/mnras/stx1777}, \href
  {https://ui.adsabs.harvard.edu/abs/2017MNRAS.471.3563D} {471, 3563}

\bibitem[\protect\citeauthoryear{Foreman-Mackey}{Foreman-Mackey}{2016}]{Foreman-Mackey2016}
Foreman-Mackey D.,  2016, \mn@doi [Journal of Open Source Software]
  {10.21105/joss.00024}, 1, 24

\bibitem[\protect\citeauthoryear{{Goncharov} et~al.,}{{Goncharov}
  et~al.}{2021}]{2021+Goncharov}
{Goncharov} B.,  et~al., 2021, \mn@doi [\apjl] {10.3847/2041-8213/ac17f4},
  \href {https://ui.adsabs.harvard.edu/abs/2021ApJ...917L..19G} {917, L19}

\bibitem[\protect\citeauthoryear{Gupta, Rickett  \& Lyne}{Gupta
  et~al.}{1994}]{gupta1994refractive}
Gupta Y.,  Rickett B.~J.,   Lyne A.~G.,  1994, Monthly Notices of the Royal
  Astronomical Society, 269, 1035

\bibitem[\protect\citeauthoryear{{Harvey-Smith}, {Madsen}  \&
  {Gaensler}}{{Harvey-Smith} et~al.}{2011}]{harvey+2011}
{Harvey-Smith} L.,  {Madsen} G.~J.,   {Gaensler} B.~M.,  2011, \mn@doi [\apj]
  {10.1088/0004-637X/736/2/83}, \href
  {https://ui.adsabs.harvard.edu/abs/2011ApJ...736...83H} {736, 83}

\bibitem[\protect\citeauthoryear{{Hellings} \& {Downs}}{{Hellings} \&
  {Downs}}{1983}]{hellings+83}
{Hellings} R.~W.,  {Downs} G.~S.,  1983, \mn@doi [\apjl] {10.1086/183954},
  \href {https://ui.adsabs.harvard.edu/abs/1983ApJ...265L..39H} {265, L39}

\bibitem[\protect\citeauthoryear{{Hemberger} \& {Stinebring}}{{Hemberger} \&
  {Stinebring}}{2008}]{hemberger+2008}
{Hemberger} D.~A.,  {Stinebring} D.~R.,  2008, \mn@doi [\apjl]
  {10.1086/528985}, \href
  {https://ui.adsabs.harvard.edu/abs/2008ApJ...674L..37H} {674, L37}

\bibitem[\protect\citeauthoryear{Hobbs et~al.,}{Hobbs
  et~al.}{2010}]{hobbs2010international}
Hobbs G.,  et~al., 2010, Classical and Quantum Gravity, 27, 084013

\bibitem[\protect\citeauthoryear{{Hotan}, {van Straten}  \&
  {Manchester}}{{Hotan} et~al.}{2004}]{Hotan+2004}
{Hotan} A.~W.,  {van Straten} W.,   {Manchester} R.~N.,  2004, \mn@doi [\pasa]
  {10.1071/AS04022}, \href
  {https://ui.adsabs.harvard.edu/abs/2004PASA...21..302H} {21, 302}

\bibitem[\protect\citeauthoryear{Kramer et~al.,}{Kramer
  et~al.}{2006}]{kramer2006tests}
Kramer M.,  et~al., 2006, Science, 314, 97

\bibitem[\protect\citeauthoryear{{Liu} et~al.,}{{Liu}
  et~al.}{2016}]{2016MNRAS.463.3239L}
{Liu} K.,  et~al., 2016, \mn@doi [\mnras] {10.1093/mnras/stw2223}, \href
  {https://ui.adsabs.harvard.edu/abs/2016MNRAS.463.3239L} {463, 3239}

\bibitem[\protect\citeauthoryear{{Lyne}}{{Lyne}}{1984}]{lyne84}
{Lyne} A.~G.,  1984, \mn@doi [\nat] {10.1038/310300a0}, \href
  {https://ui.adsabs.harvard.edu/abs/1984Natur.310..300L} {310, 300}

\bibitem[\protect\citeauthoryear{{Main} et~al.,}{{Main}
  et~al.}{2020}]{main+2020}
{Main} R.~A.,  et~al., 2020, \mn@doi [\mnras] {10.1093/mnras/staa2955}, \href
  {https://ui.adsabs.harvard.edu/abs/2020MNRAS.499.1468M} {499, 1468}

\bibitem[\protect\citeauthoryear{{Main}, {Lin}, {van Kerkwijk}, {Pen},
  {Rudnitskii}, {Popov}, {Soglasnov}  \& {Lyutikov}}{{Main}
  et~al.}{2021}]{main+21}
{Main} R.,  {Lin} R.,  {van Kerkwijk} M.~H.,  {Pen} U.-L.,  {Rudnitskii} A.~G.,
   {Popov} M.~V.,  {Soglasnov} V.~A.,   {Lyutikov} M.,  2021, \mn@doi [\apj]
  {10.3847/1538-4357/ac01c6}, \href
  {https://ui.adsabs.harvard.edu/abs/2021ApJ...915...65M} {915, 65}

\bibitem[\protect\citeauthoryear{{Maitia}, {Lestrade}  \& {Cognard}}{{Maitia}
  et~al.}{2003}]{maitia+03}
{Maitia} V.,  {Lestrade} J.~F.,   {Cognard} I.,  2003, \mn@doi [\apj]
  {10.1086/344816}, \href
  {https://ui.adsabs.harvard.edu/abs/2003ApJ...582..972M} {582, 972}

\bibitem[\protect\citeauthoryear{Manchester et~al.,}{Manchester
  et~al.}{2013}]{manchester2013parkes}
Manchester R.,  et~al., 2013, Publications of the Astronomical Society of
  Australia, 30

\bibitem[\protect\citeauthoryear{{McKee} et~al.,}{{McKee}
  et~al.}{2019}]{2019MNRAS.483.4784M}
{McKee} J.~W.,  et~al., 2019, \mn@doi [\mnras] {10.1093/mnras/sty3058}, \href
  {https://ui.adsabs.harvard.edu/abs/2019MNRAS.483.4784M} {483, 4784}

\bibitem[\protect\citeauthoryear{{Newville}, {Stensitzki}, {Allen}  \&
  {Ingargiola}}{{Newville} et~al.}{2014}]{2014zndo.....11813N}
{Newville} M.,  {Stensitzki} T.,  {Allen} D.~B.,   {Ingargiola} A.,  2014,
  {LMFIT: Non-Linear Least-Square Minimization and Curve-Fitting for Python},
  \mn@doi{10.5281/zenodo.11813}

\bibitem[\protect\citeauthoryear{{Ocker}, {Cordes}  \& {Chatterjee}}{{Ocker}
  et~al.}{2020}]{ocker+20}
{Ocker} S.~K.,  {Cordes} J.~M.,   {Chatterjee} S.,  2020, \mn@doi [\apj]
  {10.3847/1538-4357/ab98f9}, \href
  {https://ui.adsabs.harvard.edu/abs/2020ApJ...897..124O} {897, 124}

\bibitem[\protect\citeauthoryear{{Reardon} et~al.,}{{Reardon}
  et~al.}{2016}]{reardon+2016}
{Reardon} D.~J.,  et~al., 2016, \mn@doi [\mnras] {10.1093/mnras/stv2395}, \href
  {https://ui.adsabs.harvard.edu/abs/2016MNRAS.455.1751R} {455, 1751}

\bibitem[\protect\citeauthoryear{{Reardon}, {Coles}, {Hobbs}, {Ord}, {Kerr},
  {Bailes}, {Bhat}  \& {Venkatraman Krishnan}}{{Reardon}
  et~al.}{2019}]{reardon+19}
{Reardon} D.~J.,  {Coles} W.~A.,  {Hobbs} G.,  {Ord} S.,  {Kerr} M.,  {Bailes}
  M.,  {Bhat} N.~D.~R.,   {Venkatraman Krishnan} V.,  2019, \mn@doi [\mnras]
  {10.1093/mnras/stz643}, \href
  {https://ui.adsabs.harvard.edu/abs/2019MNRAS.485.4389R} {485, 4389}

\bibitem[\protect\citeauthoryear{{Reardon} et~al.,}{{Reardon}
  et~al.}{2020}]{reardon+2020}
{Reardon} D.~J.,  et~al., 2020, \mn@doi [\apj] {10.3847/1538-4357/abbd40},
  \href {https://ui.adsabs.harvard.edu/abs/2020ApJ...904..104R} {904, 104}

\bibitem[\protect\citeauthoryear{{Reardon} et~al.,}{{Reardon}
  et~al.}{2021}]{reardon+21}
{Reardon} D.~J.,  et~al., 2021, \mn@doi [\mnras] {10.1093/mnras/stab1990},
  \href {https://ui.adsabs.harvard.edu/abs/2021MNRAS.507.2137R} {507, 2137}

\bibitem[\protect\citeauthoryear{{Rickett} et~al.,}{{Rickett}
  et~al.}{2014}]{rickett+14}
{Rickett} B.~J.,  et~al., 2014, \mn@doi [\apj] {10.1088/0004-637X/787/2/161},
  \href {https://ui.adsabs.harvard.edu/abs/2014ApJ...787..161R} {787, 161}

\bibitem[\protect\citeauthoryear{{Shannon} et~al.,}{{Shannon}
  et~al.}{2016}]{shannon+2016}
{Shannon} R.~M.,  et~al., 2016, \mn@doi [\apjl] {10.3847/2041-8205/828/1/L1},
  \href {https://ui.adsabs.harvard.edu/abs/2016ApJ...828L...1S} {828, L1}

\bibitem[\protect\citeauthoryear{{Smits} et~al.,}{{Smits}
  et~al.}{2017}]{2017A&C....19...66S}
{Smits} R.,  et~al., 2017, \mn@doi [Astronomy and Computing]
  {10.1016/j.ascom.2017.02.002}, \href
  {https://ui.adsabs.harvard.edu/abs/2017A&C....19...66S} {19, 66}

\bibitem[\protect\citeauthoryear{{Sprenger}, {Wucknitz}, {Main}, {Baker}  \&
  {Brisken}}{{Sprenger} et~al.}{2021}]{sprenger+2021}
{Sprenger} T.,  {Wucknitz} O.,  {Main} R.,  {Baker} D.,   {Brisken} W.,  2021,
  \mn@doi [\mnras] {10.1093/mnras/staa3353}, \href
  {https://ui.adsabs.harvard.edu/abs/2021MNRAS.500.1114S} {500, 1114}

\bibitem[\protect\citeauthoryear{Stinebring, McLaughlin, Cordes, Becker,
  Goodman, Kramer, Sheckard  \& Smith}{Stinebring
  et~al.}{2001}]{stinebring2001faint}
Stinebring D.,  McLaughlin M.,  Cordes J.,  Becker K.,  Goodman J.~E.,  Kramer
  M.,  Sheckard J.,   Smith C.,  2001, The Astrophysical Journal Letters, 549,
  L97

\bibitem[\protect\citeauthoryear{{Stinebring}, {Rickett}  \&
  {Ocker}}{{Stinebring} et~al.}{2019}]{2019+Stinebring}
{Stinebring} D.~R.,  {Rickett} B.~J.,   {Ocker} S.~K.,  2019, \mn@doi [\apj]
  {10.3847/1538-4357/aaef80}, \href
  {https://ui.adsabs.harvard.edu/abs/2019ApJ...870...82S} {870, 82}

\bibitem[\protect\citeauthoryear{Taylor, Fowler  \& McCulloch}{Taylor
  et~al.}{1979}]{taylor1979measurements}
Taylor J.~H.,  Fowler L.,   McCulloch P.,  1979, Nature, 277, 437

\bibitem[\protect\citeauthoryear{Verbiest et~al.,}{Verbiest
  et~al.}{2016}]{verbiest2016international}
Verbiest J.,  et~al., 2016, Monthly Notices of the Royal Astronomical Society,
  458, 1267

\bibitem[\protect\citeauthoryear{Walker, Melrose, Stinebring  \& Zhang}{Walker
  et~al.}{2004}]{walker2004interpretation}
Walker M.~A.,  Melrose D.~B.,  Stinebring D.,   Zhang C.,  2004, Monthly
  Notices of the Royal Astronomical Society, 354, 43

\bibitem[\protect\citeauthoryear{{Yao} et~al.,}{{Yao} et~al.}{2021}]{yao+21}
{Yao} J.,  et~al., 2021, \mn@doi [Nature Astronomy]
  {10.1038/s41550-021-01360-w}, \href
  {https://ui.adsabs.harvard.edu/abs/2021NatAs...5..788Y} {5, 788}

\bibitem[\protect\citeauthoryear{van Haasteren et~al.,}{van Haasteren
  et~al.}{2011}]{van2011placing}
van Haasteren R.,  et~al., 2011, Monthly Notices of the Royal Astronomical
  Society, 414, 3117

\bibitem[\protect\citeauthoryear{{van Straten} \& {Bailes}}{{van Straten} \&
  {Bailes}}{2011}]{2011PASA...28....1V}
{van Straten} W.,  {Bailes} M.,  2011, \mn@doi [\pasa] {10.1071/AS10021}, \href
  {https://ui.adsabs.harvard.edu/abs/2011PASA...28....1V} {28, 1}

\bibitem[\protect\citeauthoryear{{van Straten}, {Demorest}  \& {Oslowski}}{{van
  Straten} et~al.}{2012}]{2012AR&T....9..237V}
{van Straten} W.,  {Demorest} P.,   {Oslowski} S.,  2012, Astronomical Research
  and Technology, \href {https://ui.adsabs.harvard.edu/abs/2012AR&T....9..237V}
  {9, 237}

\makeatother
\end{thebibliography}

\clearpage

\appendix

\onecolumn

\section{Simulations of different Screen Geometries}
\label{sec:screengeometry}
In Section \ref{sec:arc_qualitative} and Figure \ref{fig:panorama}, we note that the power distribution we see cannot be reproduced with a single 1-D screen or a perfectly isotropic screen, and adopt the possibility of multiple screens, or an elongated 2D screen. Here, we performed simulations to illustrate the different models, with the same simulation code used in \citet{baker+21}.  

The simulation uses a set of image positions along a thin screen, treating each image as a stationary phase point with a random amplitude and phase at each point. The combination of dispersive and geometric delays remains constant along the screen at a reference frequency. For a grid of time and frequency values, the electric field at the observer is computed as the coherent summation of each point,
\begin{equation}
E(t_{i}, f_{j}) = \sum_{k} \mu_{k} e^{i\phi_{k}}.
\end{equation}
The relative geometric delays of the images change over time, producing the time-variable electric field. The dynamic spectrum is then calculated from the amplitude squared of the electric field, and the secondary spectrum as the squared modulus of the 2D FFT of the dynamic spectrum.

\begin{figure*}
\begin{minipage}{.78\textwidth}
\includegraphics[width=0.99\textwidth,trim=4.0cm 0.0cm 4.5cm 0.8cm, clip=true]{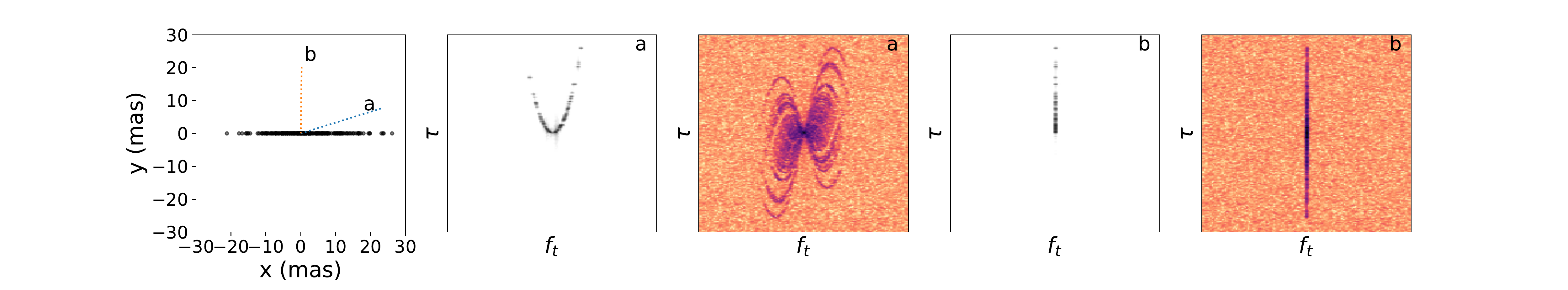}\\
\includegraphics[width=0.99\textwidth,trim=4.0cm 0.0cm 4.5cm 0.8cm, clip=true]{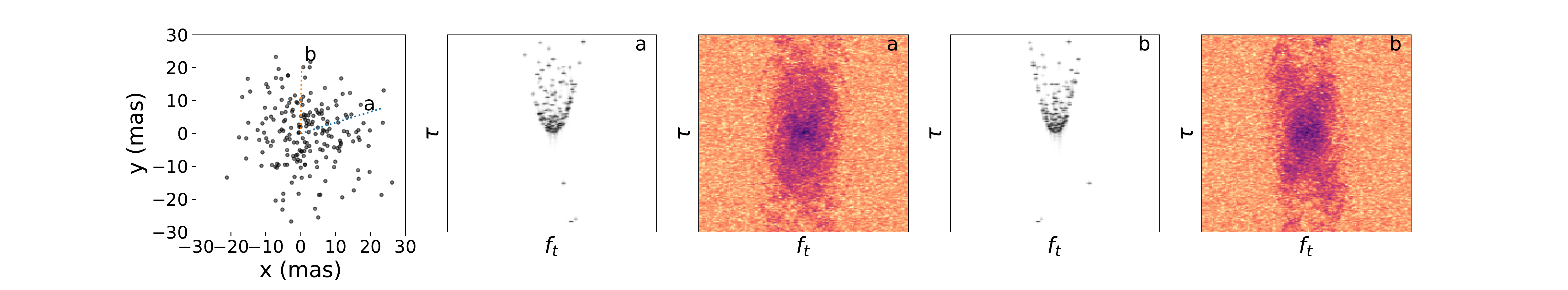}\\
\includegraphics[width=0.99\textwidth,trim=4.0cm 0.0cm 4.5cm 0.8cm, clip=true]{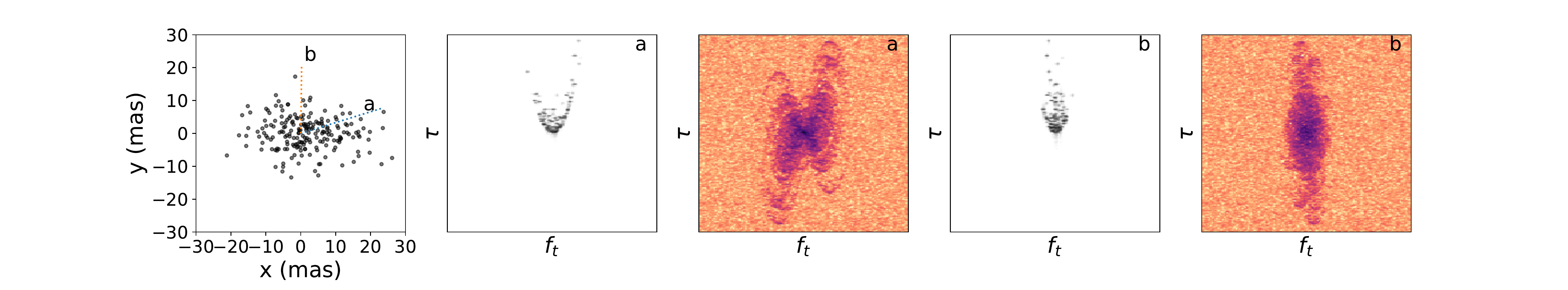}\\
\includegraphics[width=0.99\textwidth,trim=4.0cm 0.0cm 4.5cm 0.8cm, clip=true]{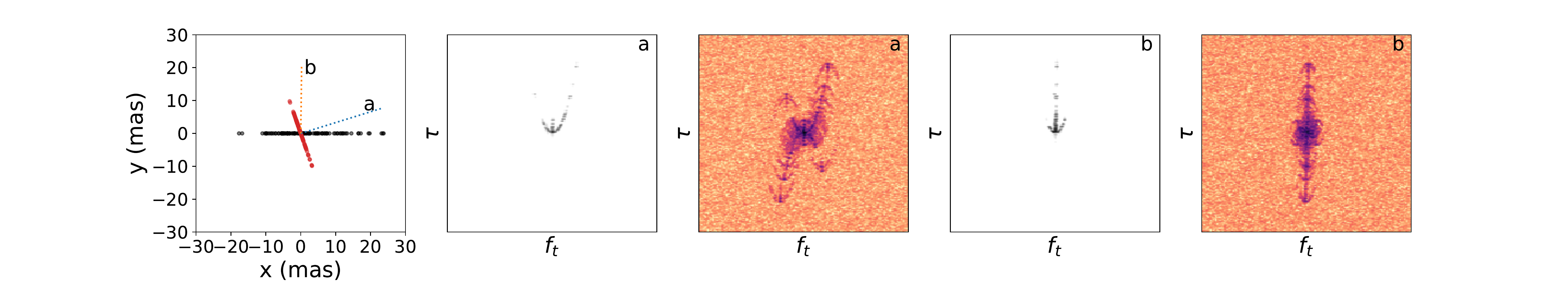}\\
\end{minipage}
\caption{Simulations of different screen geometries, seen at two different effective velocity vectors. Greyscale images show the wavefield $|E(f_{t}, \tau)|^{2}$, while the coloured images show the secondary spectra $|I(f_{t}, \tau)|^{2}$ Top: a perfectly 1D screen, Second row: a statistically isotropic 2D screen, Third row: a 2D screen with a 2:1 axial ratio, Bottom: two misaligned 1D screens.  Details of the simulations are given in Appendix \ref{sec:screengeometry}. Qualitatively, neither the perfectly 1D screen, or the fully isotropic screen fit the distribution of power observed for \psr{} throughout the year, with the 1D screen collapsing too thin when $\textbf{v}_{\rm{eff}} \rightarrow 0$, and the isotropic screen never showing clear arcs. An anisotropic 2D screen, or two 1D screens can produce clear primary arcs, and a dominant lower curvature at low time delays depending on the orientation of $\textbf{v}_{\rm{eff}}$}
\label{fig:sim}
\end{figure*}

We perform these simulations for the four geometries, a 1D screen, a statistically isotropic screen, a 2D screen with a 2:1 axial ratio, and two misaligned 1D screens at different distances. In each case, we simulate 200 random stationary phase points. In Figure \ref{fig:sim} we show the image distribution of each simulation, and secondary spectra corresponding to two different values of {\veff}. For the case of two screens, all pairs of images are allowed to interfere, while multiply-deflected paths are not considered.

Qualitatively, a 1D screen collapses to the $f_{D}=0$ when {\veff} $\rightarrow 0$, inconsistent with what we see for \psr{}. The other two models can qualitatively reproduce our results; when the velocity is aligned with the dominant axis (either the elongated axis in 2D, or the dominant screen with two 1D screens), then we can observe a dominant scintillation arc, which is only slightly smeared.  When the velocity is misaligned with the dominant axis, the curvature at low time delays is dominated by the velocity projected along the second axis (proportional to {\veff} for a 2D screen, $|v_{\rm eff}|$ parallel to the second screen for two screens), while the power at high delays is caused by images along the dominant axis interfering with all other images.

\clearpage

\section{Posterior Distributions}
\label{sec:posteriors}
Figures B1 and B2 show the posterior distributions between different parameters. In these figures blue lines indicate mean values and contours show $1\sigma$, $2\sigma$ and $3\sigma$ confidence levels. Figure B1 shows the distributions resulting from the isotropic model discussed in Section {\ref{sec:isotropic}} and Figure B2 shows the distributions resulting from the anisotropic two-screen model discussed in Section {\ref{sec:anisotropic}}. In both of these figures we observe high covariances between pulsar's and screen's distance and velocity.

\begin{figure*}
\centering
\includegraphics[width=0.7\textwidth,trim=0.0cm 0cm 0cm 0.0cm, clip=true]{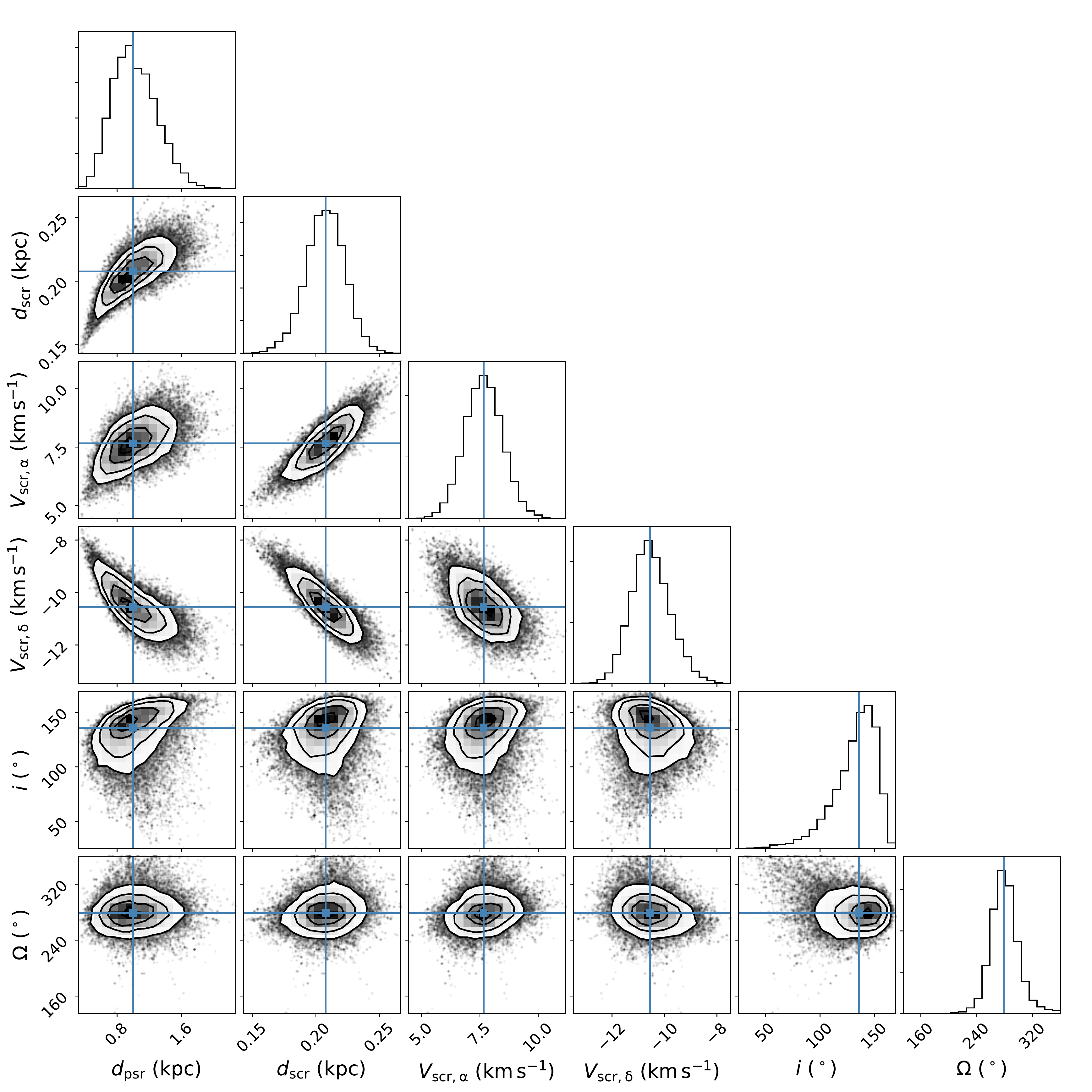}\\
\caption{Posterior probability distributions for the orbital and scintillation parameters of J1643-1224, obtained using an isotropic model fit and a MCMC sampler. The blue lines indicate the mean values and the contour show $1\sigma$, $2\sigma$ and $3\sigma$ confidence levels. The plot shows significant covariances between the parameter pairs ($\rm{d_{scr}}$, $\rm{{V}_{scr,\alpha}}$), ($\rm{d_{scr}}$, $\rm{{V}_{scr,\delta }}$), (d, $\rm{{V}_{scr,\delta }}$), and (d, $\rm{d_{scr} }$) with correlation coefficients of 0.83, -0.86, -0.78 and 0.76 respectively. This figure was created with corner.py \protect\citep{Foreman-Mackey2016}}.
\label{fig:mcmc2D}
\end{figure*}

\begin{figure*}
\centering
\includegraphics[width=1.0\textwidth,trim=0.0cm 0cm 0cm 0.0cm, clip=true]{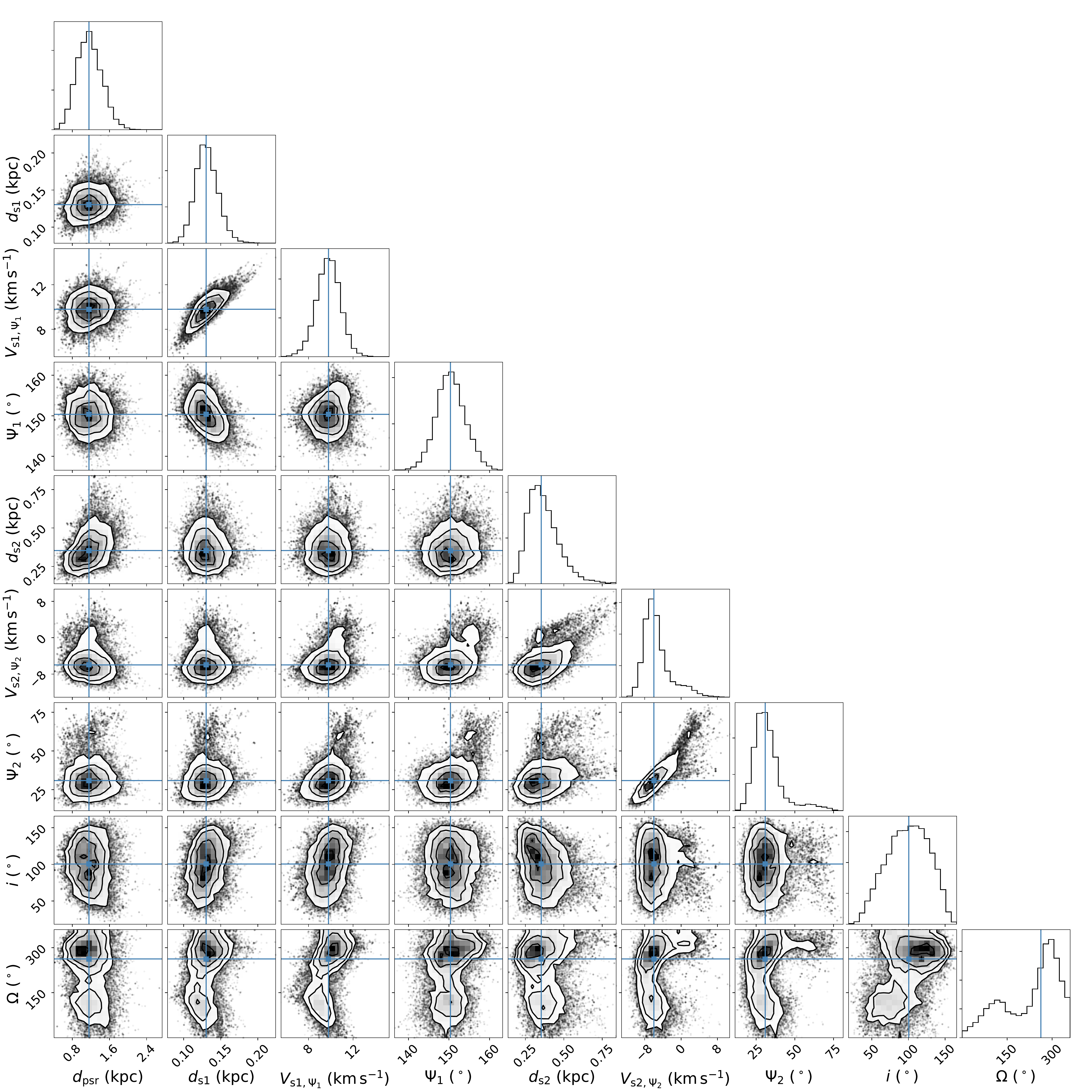}\\
\caption{Posterior probability distributions for the orbital and scintillation parameters of J1643-1224, obtained using an anisotropic two-screen model fit and a MCMC sampler. The blue lines indicate the mean values and the contour show $1\sigma$, $2\sigma$ and $3\sigma$ confidence levels. There are two local maxima in $i$ and $\Omega$, so this model cannot unambiguously determine the sense of orbit. The plot shows significant covariances between the parameter pairs ($\rm{V_{{s2},\Psi_{2}}}$,$\Psi_{2}$) and ($\rm{d_{s1}}$, $\rm{V_{{s1},\Psi_{1}}}$) with correlation coefficients of 0.91, and 0.79 respectively. This figure was created with corner.py \protect\citep{Foreman-Mackey2016}.}

\label{fig:mcmctwoscreen}
\end{figure*}

\bsp	
\label{lastpage}
\end{document}